\input{aipcheck}

\documentclass[
  ,final            
  ]
  {aipproc}

\usepackage{graphicx}
\usepackage{amsmath}
\usepackage{txfonts}
\usepackage{natbib}
\usepackage{subfigure}

\newcommand{\eh}[1]{\,\mathrm{#1}}

\newcommand{\dg}{^{\circ}}

\newcommand{\mr}[1]{\mathrm{#1}}

\renewcommand{\epsilon}{\varepsilon}

\newcommand{\tin}[1]{_{\mr{#1}}}

\newcommand{\w}{\omega}
\newcommand{\beq}{\begin{equation}}
\newcommand{\eeq}{\end{equation}}

\newcommand{\beqa}{\begin{eqnarray}}
\newcommand{\eeqa}{\end{eqnarray}}

\layoutstyle{6x9}
\renewcommand\XFMtitleblock{%
  \XFMtitle
  \let\XFMoldpar\par
  \def\par{\XFMoldpar\def\par{\space
             for the MAGIC Collaboration\XFMoldpar}}%
   \XFMauthors
   \let\par\XFMoldpar
   \XFMaddresses
   \XFMabstract
   \vspace{5pt}%
   \XFMkeywords
   \XFMclassification
 }

\begin{document}

\title{Application of a generalized likelihood ratio\\ test statistic to MAGIC
data}


\author{S. Klepser}{
  address={IFAE, Edifici Cn., Campus UAB, E-08193 Bellaterra, Spain}
  ,altaddress={Deutsches Elektronen-Synchrotron (DESY), D-15738 Zeuthen,
Germany}
}

\author{J. Krause}{
  address={Max-Planck-Institut f\"ur Physik, D-80805 M\"unchen, Germany}
}

\author{J. Sitarek}{
  address={IFAE, Edifici Cn., Campus UAB, E-08193 Bellaterra, Spain}
}


\begin{abstract}
The commonly used detection test statistic for Cherenkov telescope
data is Li \& Ma (1983), Eq. 17. It evaluates the compatibility of event counts in an
on-source region with those in a representative off-region. It does not
exploit the typically known gamma-ray point spread function (PSF) of a system, and in practice its
application
requires either assumptions on the symmetry of the acceptance across the field
of view, or
Monte Carlo simulations. MAGIC has an azimuth-dependent, asymmetric acceptance which required
a careful review of detection statistics. Besides an adapted Li \& Ma based
technique, the recently presented generalized LRT statistic of \cite{lrt} is now in use. It is more flexible, more sensitive and less
systematics-affected, because it is highly customized for multi-pointing
Cherenkov telescope data with a known PSF.
We present the application of this new method
to archival MAGIC data and compare it to the other, Li\&Ma-based method.
%
%
%
\end{abstract}

\maketitle



  MAGIC \cite{magicstereoperformance} is a Cherenkov telescope
  system that comprises two large
telescopes and cameras with fields of view (FOV) of
$3.5\dg$ in diameter. The sensitive trigger area diameters are about $2.0\dg$ and
$2.4\dg$. This comparably small area leads to a stereoscopic
  gamma-ray FOV that is not circularly symmetric and whose shape rotates with the
  azimuth angle of observation. Uniformity-based background
  estimation methods like in \cite{hessskymapping} cannot reliably
be applied, neither for skymapping nor source detection in general.

%


\section{Source detection with MAGIC}

\begin{figure}[t]
  \begin{minipage}{4.2cm}
  \includegraphics[height=.15\textheight]{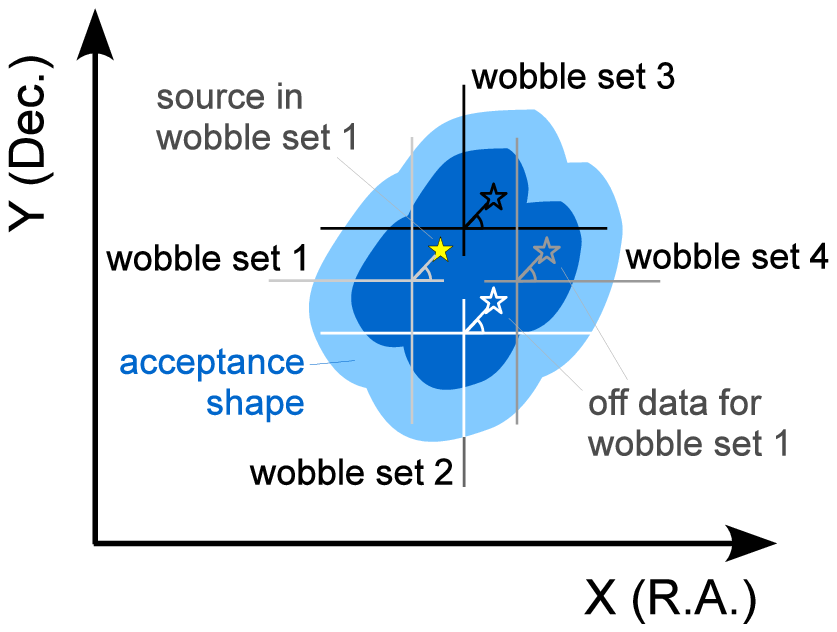}
  \end{minipage}
\centering
$\begin{array}{c@{\hspace{0.001cm}}c}
  \multicolumn{1}{l}{\mbox{\textbf{} }} & \multicolumn{1}{l}{\mbox{\textbf{}
}} \\
  \mbox{\footnotesize{\textbf{\textsf{>300 GeV}}}} &
\mbox{\footnotesize{\textbf{\textsf{>100 GeV}}}} \\
  \resizebox{0.38\hsize}{!}{\includegraphics{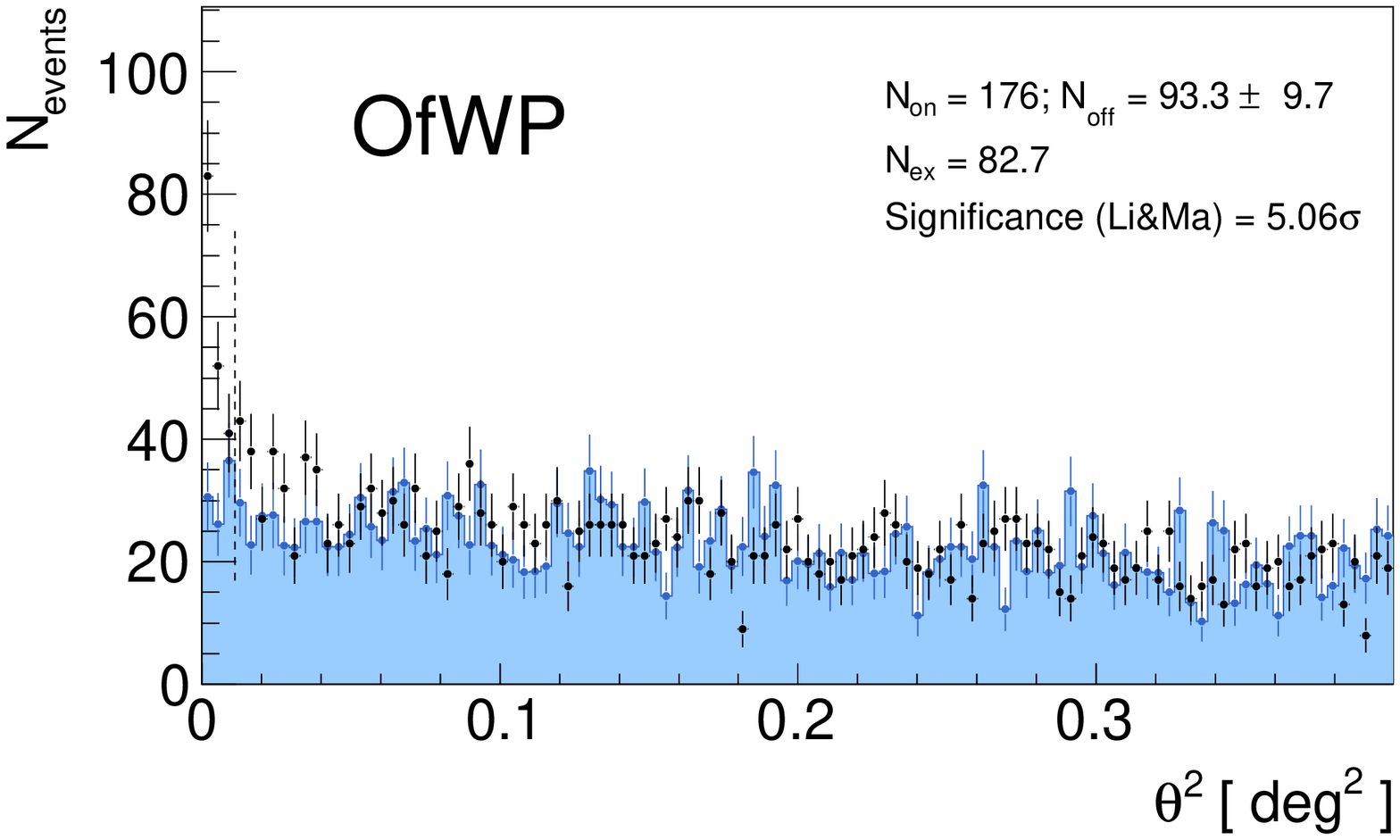}}&
  \resizebox{0.38\hsize}{!}{\includegraphics{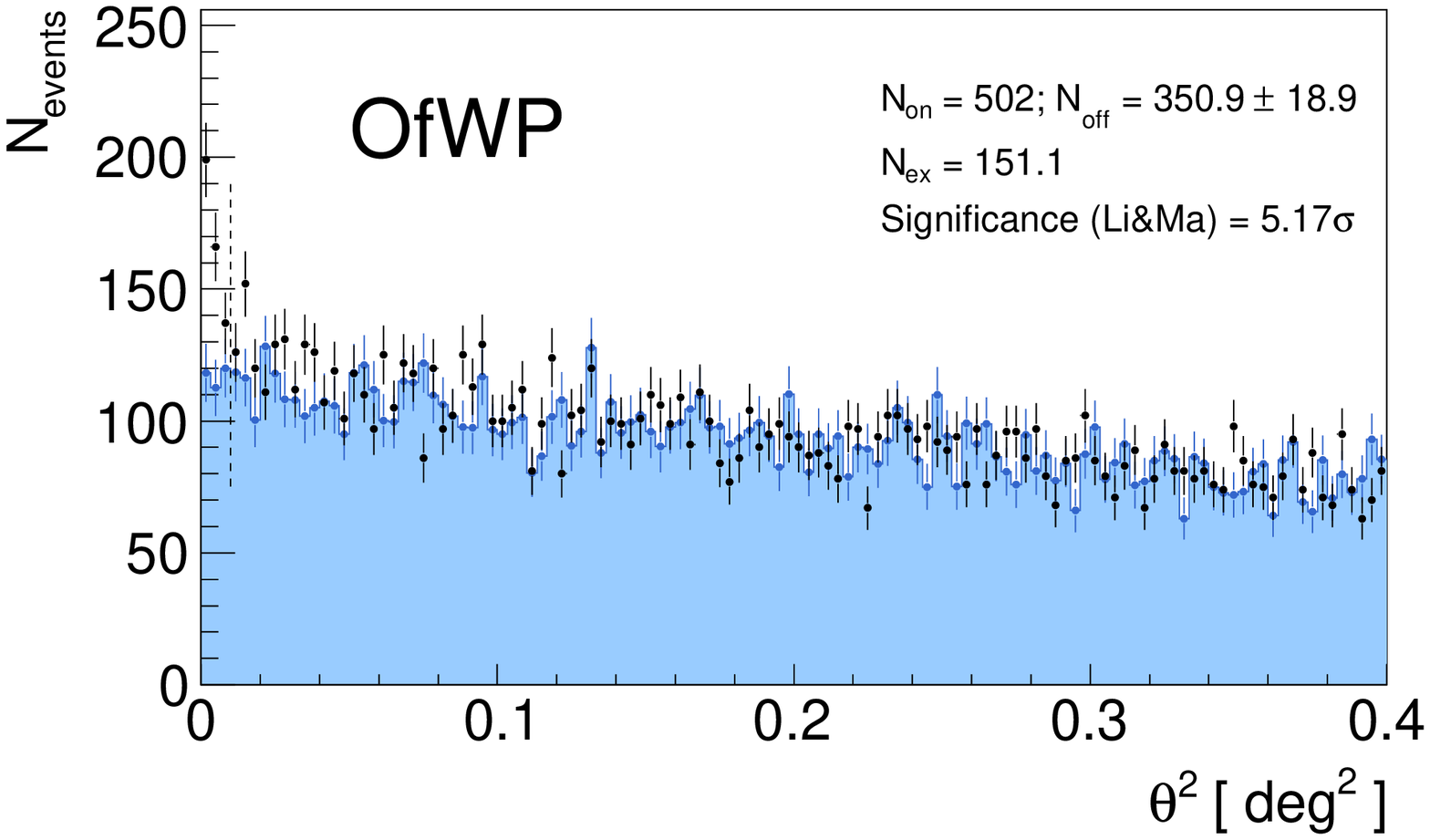}} \\
  \resizebox{0.38\hsize}{!}{\includegraphics{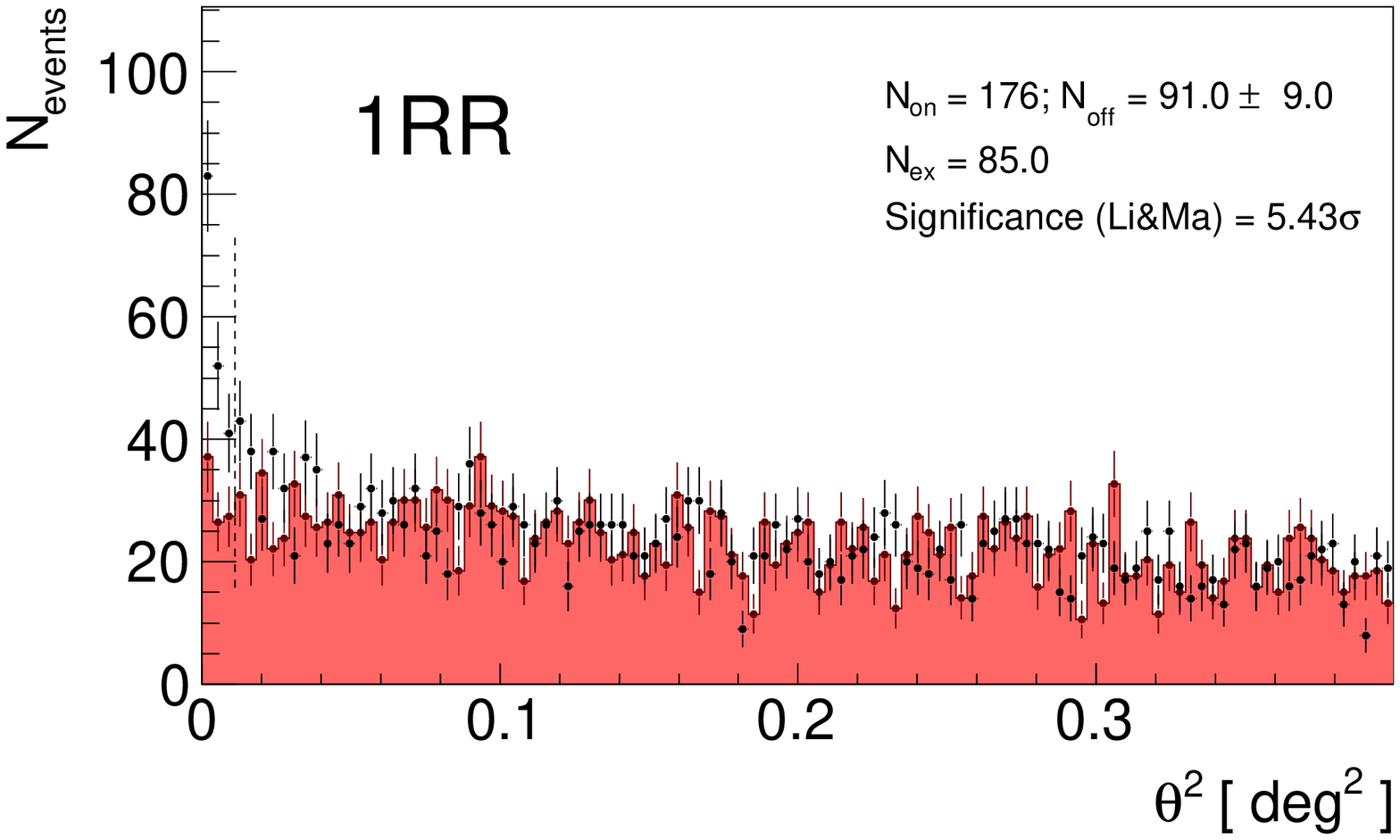}}&
  \resizebox{0.38\hsize}{!}{\includegraphics{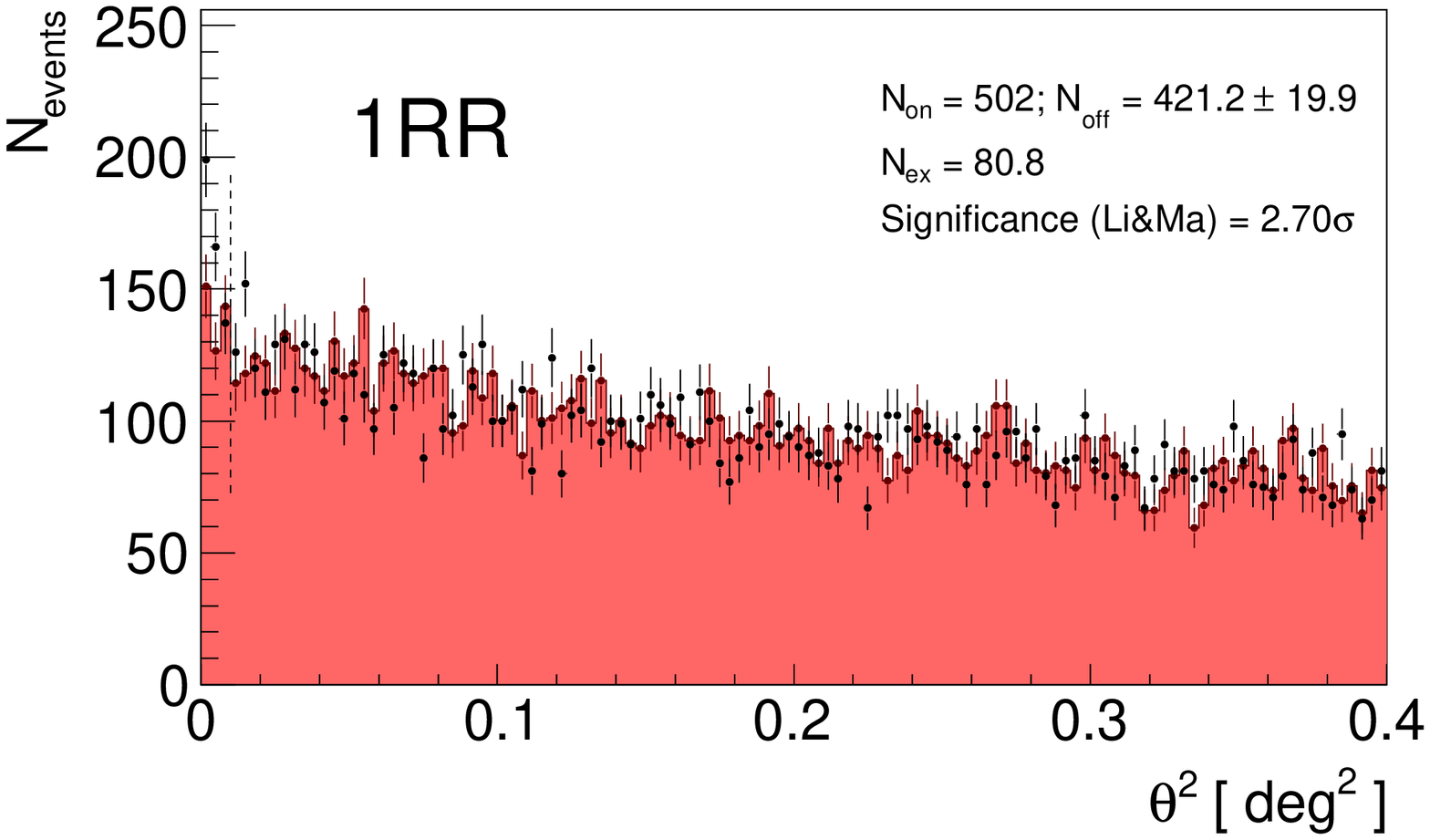}} \\
\end{array}$
\caption{Left: The Off-from-Wobble-Partner scheme in a 4-wobble observation. Right:
Distributions of squared distance between gamma direction
and source position ($\theta^2$) for high- and low-energy cuts, using OfWP
(top row) and 1RR (bottom row) methods, and the Li \& Ma, Eq.~17 test statistic.}
\label{fig:th2plots}
\end{figure}

What has become the standard way of source detection with MAGIC, at least at low
energies, is to observe several positions around the target source coordinate
(wobble mode), and apply the so-called
\textit{Off-from-Wobble-Partner} (OfWP) method.
In this scheme, the on-source event count of
a given wobble set is compared to the off-counts extracted from one or more
other wobble sets, using the same focal plane coordinate as the on-data
(Fig.~\ref{fig:th2plots}, left). The
off-counts have to be scaled to the on-data, either by effective observation time or background rate. The
data of different wobble sets are summed up, and the test statistic of \cite{lima}, Eq.~17 can be applied,
assuming an effective $\alpha$-parameter of
$\alpha\tin{eff}=\sum_{\w}\alpha_{\w}^2 N_{\w}^{\mr{off}}/\sum_{\w}\alpha_{\w}
N_{\w}^{\mr{off}}$,
where $\alpha_{\w}$ are the normalization factors of the wobble sets $\w$. The
validity of this method is a plausible approximation if all $\alpha_{\w}$ are
similar, but is uncertain if not. This happens in unbalanced wobble
data, which is particularly likely to happen if the data are binned in
azimuth and randomly sized portions of events can occur.
A second drawback is that in Li \& Ma,
Eq.~17, the gamma-ray PSF is not exploited, although it is a well-known
performance parameter that can help to distinguish a signal from a background fluctuation.
Finally, the OfWP and "reflected regions" \cite{hessskymapping} methods are only correct if the
off-regions are well-separated from potential sources, and do not overlap with
each other. This often requires manual case-by-case
consideration, which is not optimal for automatic 
procedures.


The likelihood ratio test statistic (LRT) of \cite{lrt}, Eq.~16 is a
generalized version of the Li \& Ma formula. It is customized to multi-wobble
Cherenkov telescope observations. The generalized LRT accommodates
that the data are taken in various wobble sets, throughout several "operating
conditions" (azimuth or zenith angle, weather, ...), and that a signal, if present, will
appear with a possibly known PSF shape. The calculation involves the numerical
determination of a \textit{relative excess parameter} $\phi\tin{sup}$, which
describes the amplitude of the signal relative to the background.
%
%

\section{Test data set and standard analysis}

We use a data subset that comprises two wobble sets in which the radio galaxy
IC~310 can
marginally be detected at off-axis angles of $0.25\dg$ and $1.0\dg$ from the observed coordinates (see
\cite{ic310} for the physics discussion and analysis of the full data set).
We tested the detection significance for two energy cuts ($>100\eh{GeV}$,
$>300\eh{GeV}$), using pre-defined, multi-purpose event selection cuts.
On the right side of Fig.~\ref{fig:th2plots} is a comparison of detection plots using OfWP and one reflected region (1RR, see \cite{hessskymapping}).
Clearly, OfWP and 1RR are consistent at high energies, but show a significant
difference at lower energies. This behaviour is not exceptional, and studies done with off-data show that
in those cases, OfWP deals better with the inhomogeneous background acceptance
and is generally more reliable.

\section{Application of the new test statistic}

The generalized LRT is implemented in two steps. In the first step, which
is part of the MAGIC analysis package MARS \cite{magicstereoperformance},
the gamma-ray events, on-time and observation coordinates are filled into MARS-independent
ROOT\footnote{http://root.cern.ch/} histograms and
containers. The data are binned in ranges of azimuth angle to account for the
changing acceptance shape. In the second step, the on-source significance and
significance skymap are calculated. This latter part uses only ROOT routines
and is easily applicable
to data of other experiments as well.
The whole procedure is automatic, i.e. any number of data sets
are combined without any manual effort. The determination of
$\phi\tin{sup}$ is done using the method of \citet{riddersmethod}.

\subsection{Trials and self-consistency check}

The skymaps shown in Fig.~\ref{fig:skymaps} are calculated in a fine grid of sky
coordinates, leading to correlated significance values.
Therefore, the significance \textit{distributions} are calculated using a much rougher
grid with steps of $g = \sqrt{2\,\pi}\sigma$, where $\sigma$ is the gamma-ray
PSF. This rough scan is repeated 9 times
with offsets of $g/3$, in order not to miss a source. These 9 subscans are
conservatively taken into account as independent trials.

Instead of $\theta^2$-histograms, the generalized LRT method provides a skymap and significance
distribution. This is a much more complete consistency check since the Gaussian
nature of the significance value is verified for every detection, 
which is not
possible in a $\theta^2$-histogram detection.

\section {Results and Conclusion}

\begin{figure}
  \includegraphics[height=.35\textheight]{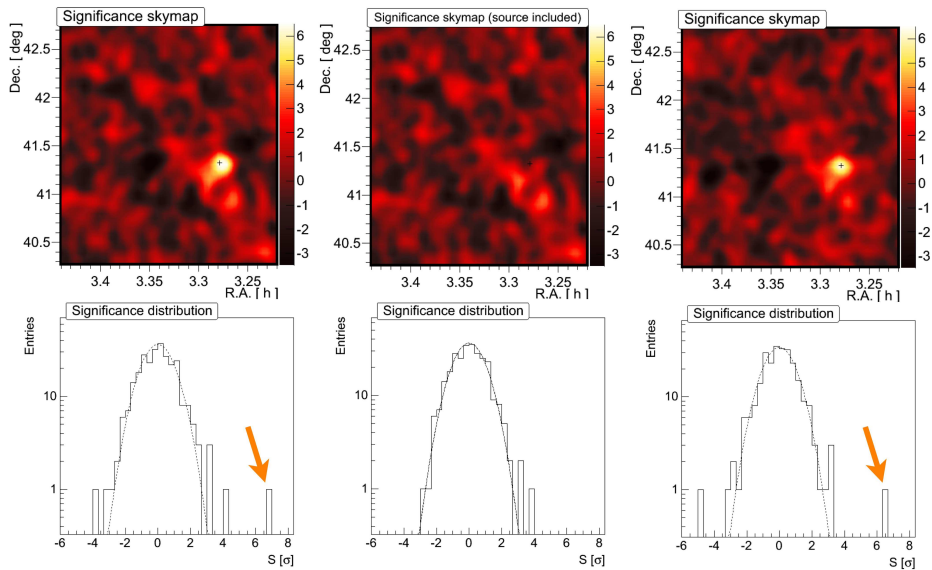}
  \caption{Significance skymap and distribution for the high-energy cut
($>300\eh{GeV}$, left), including the source into the null hypothesis
($>300\eh{GeV}$, middle), and for low energies ($>100\eh{GeV}$, right). The
distributions use a rougher sky grid than the skymaps (to avoid correlations, see text). The dotted line is a
Gaussian function ($\mu=0$, $\sigma=1$).}
  \label{fig:skymaps}
\end{figure}

The generalized LRT method was first tested using observations of sky regions without
known sources  (off-data). 
No source was found, and the significance distribution was Gaussian as
expected.
Figure~\ref{fig:skymaps} shows the significance skymaps and distributions for
high- and low-energy cuts, and also a skymap where the source is incorporated
in the null hypothesis (middle
column of Fig.~\ref{fig:skymaps}). The on-source significances are
$6.6\eh{\sigma}$ ($5.6\eh{\sigma}$) and $6.5\eh{\sigma}$ ($5.2\eh{\sigma}$)
pre-trial (post-trial), respectively. The
significances are somewhat higher than the above OfWP significances, and
no systematic inconsistency is found at low energies. The source-including skymap leads to a
Gaussian distribution with a post-trial significance of
$2.2\eh{\sigma}$ - i.e. no additional source is detected.

%
Concluding, the generalized LRT method \cite{lrt} was successfully applied to MAGIC data and, as
claimed, tends to be more sensitive than the test statistic of Li \& Ma,
Eq.~17. It was implemented as a stand-alone ROOT macro that can be applied
flexibly and without manual selection of off-regions.

%


\begin{theacknowledgments}
We would like to thank the Instituto de Astrof\'{\i}sica de
Canarias for the excellent working conditions at the
Observatorio del Roque de los Muchachos in La Palma.
The support of the German BMBF and MPG, the Italian INFN, 
the Swiss National Fund SNF, and the Spanish MICINN is 
gratefully acknowledged. This work was also supported by the CPAN
CSD2007-00042 and MultiDark
CSD2009-00064 projects of the Spanish Consolider-Ingenio 2010
programme, by grant DO02-353 of the Bulgarian NSF, by grant 127740 of 
the Academy of Finland,
by the DFG Cluster of Excellence ``Origin and Structure of the 
Universe'', by the DFG Collaborative Research Centers SFB823/C4 and SFB876/C3,
and by the Polish MNiSzW grant 745/N-HESS-MAGIC/2010/0.
\end{theacknowledgments}



\bibliographystyle{aipproc}   

\bibliography{lrt_magic}

\IfFileExists{\jobname.bbl}{}
 {\typeout{}
  \typeout{******************************************}
  \typeout{** Please run "bibtex \jobname" to optain}
  \typeout{** the bibliography and then re-run LaTeX}
  \typeout{** twice to fix the references!}
  \typeout{******************************************}
  \typeout{}
 }

\end{document}